\documentclass[a4paper]{article}
\usepackage{INTERSPEECH2020}
\usepackage{graphicx}
\usepackage{amssymb, amsmath, bm, mathtools,array}
\usepackage{textcomp}
\usepackage{hyperref}
\usepackage{verbatim}
\usepackage{lipsum} 
\usepackage{multirow}
\usepackage{siunitx}
\usepackage{diagbox}

\newcommand{\bs}[1]{\boldsymbol{#1}}

\RequirePackage{color}\definecolor{BLUE}{rgb}{0,0,0} 

\usepackage{caption}
\usepackage{subcaption}

\title{Using Cyclic Noise as the Source Signal for Neural Source-Filter-based Speech Waveform Model}
\name{Xin Wang$^1$, Junichi Yamagishi$^{1,}$$^2$}
\address{
  $^1$National Institute of Informatics, Japan, $^2$CSTR, University of Edinburgh, UK}
\email{wangxin@nii.ac.jp, jyamagis@nii.ac.jp}

\begin{document}

\maketitle
\begin{abstract}
Neural source-filter (NSF) waveform models generate speech waveforms by morphing sine-based source signals through dilated convolution in the time domain. Although the sine-based source signals help the NSF models to produce voiced sounds with specified pitch, the sine shape may constrain the generated waveform when the target voiced sounds are less periodic. In this paper, we propose a more flexible source signal called cyclic noise, a quasi-periodic noise sequence given by the convolution of a pulse train and a static random noise with a trainable decaying rate that controls the signal shape. We further propose a masked spectral loss to guide the NSF models to produce periodic voiced sounds from the cyclic noise-based source signal. 
Results from a large-scale listening test demonstrated the effectiveness of the cyclic noise and the masked spectral loss on speaker-independent NSF models in copy-synthesis experiments on the CMU ARCTIC database. 
\end{abstract}

\noindent\textbf{Index Terms}: speech synthesis, source-filter model, harmonic-plus-noise waveform model, neural network

\section{Introduction}
Neural waveform models are neural network (NN)-based generative models that produce waveforms given acoustic features, linguistic features, other types of input data. In addition to the pioneering work using feedforward networks for speech enhancement \cite{tamura1988noise,gao1996speech}, many new types of neural waveform models have been proposed recently for text-to-speech (TTS) synthesis \cite{tokuday2015directly, oord2016wavenet, prenger2018waveglow}, band extension \cite{kuleshov2017audio}, music signal generation \cite{engel17a, engel2019gansynth}, and other tasks. 
While new models such as WaveNet \cite{oord2016wavenet} and WaveGlow \cite{prenger2018waveglow} are based on the latest deep learning methods, many other models combine NNs with classical signal processing techniques such as adaptive cepstral analysis \cite{tokuday2015directly}, linear prediction coding (LPC) \cite{valin2018lpcnet}, and glottal waveform modeling \cite{juvela2019glotnet}.  
These models can efficiently produce high-quality waveforms by leveraging signal processing algorithms that cannot be easily learned by neural networks \cite{engel2020ddsp}.

For TTS, we have proposed a family of neural source-filter (NSF) waveform models \cite{wang2018neural, wangNSFall,Wang2019} that combines the dilated convolution (CONV) NN with the idea of the source-filter speech production model. The NSF models share three common components: a condition module that pre-processes the input spectral features, a source module that produces sine-based source signals from input F0, and a dilated-CONV-based filter module that converts the source signals into output waveforms. 
Compared with other neural waveform models, NSF models are easy to train and work fast for waveform generation, and our previous study showed their good performance in a Japanese TTS system with a female voice \cite{wangNSFall}.

The source signal is essential to NSF models as these NSF models produce output waveforms by gradually morphing the source signal in the time domain. 
In our previous works on TTS, we used sine-based source signals because 
their periodicity can be accurately maintained in the generated voiced sounds \cite{wangNSFall}. 
In addition to the speech waveforms, the sine-based waveform also helps the NSF models to produce high-quality music signals for woodwind, string, and brass instruments \cite{Zhao2020}. 

However, for speech waveform generation, the sine-based source signal may cause artifacts when the target sounds are less periodic, e.g., when they are creaky or breathy \cite{gordon2001phonation}.
To relax the constraint from using a specific waveform shape while maintaining the periodicity,  we propose using a cyclic noise as the source signal. We compute this cyclic noise by convolving a pulse train of the specified F0 and a random noise with a tunable decaying rate, which keeps the quasi periodicity of the source signal and allows cyclostationary randomness across periods. We also propose a masked spectral loss that encourages the NSF models to produce periodic voiced sounds from the quasi-periodic cyclic noise.  Experimental results demonstrated that the proposed cyclic noise outperformed the sine-based source signal on male speaker data while performing equally well on female speaker data. 

After we briefly review the latest NSF model in Section~\ref{seq:review}, we describe the proposed source signal and loss in Section~\ref{seq:proposed} and explain the experiment results in Section~\ref{seq:exp}. We conclude in Section~\ref{sec:conclude} with a brief summary. Although we focus on NSF models in this paper, we hope that our findings can be useful to related works that use random noises \cite{prenger2018waveglow, juvela2019gelp}, LPC residuals \cite{valin2018lpcnet}, or other periodic signals \cite{watts2019speech, ai2020neural} as the source signals .

\section{Review of hn-sinc-NSF model}
\label{seq:review}
\begin{figure}[!t]
\centering
\includegraphics[width=\columnwidth]{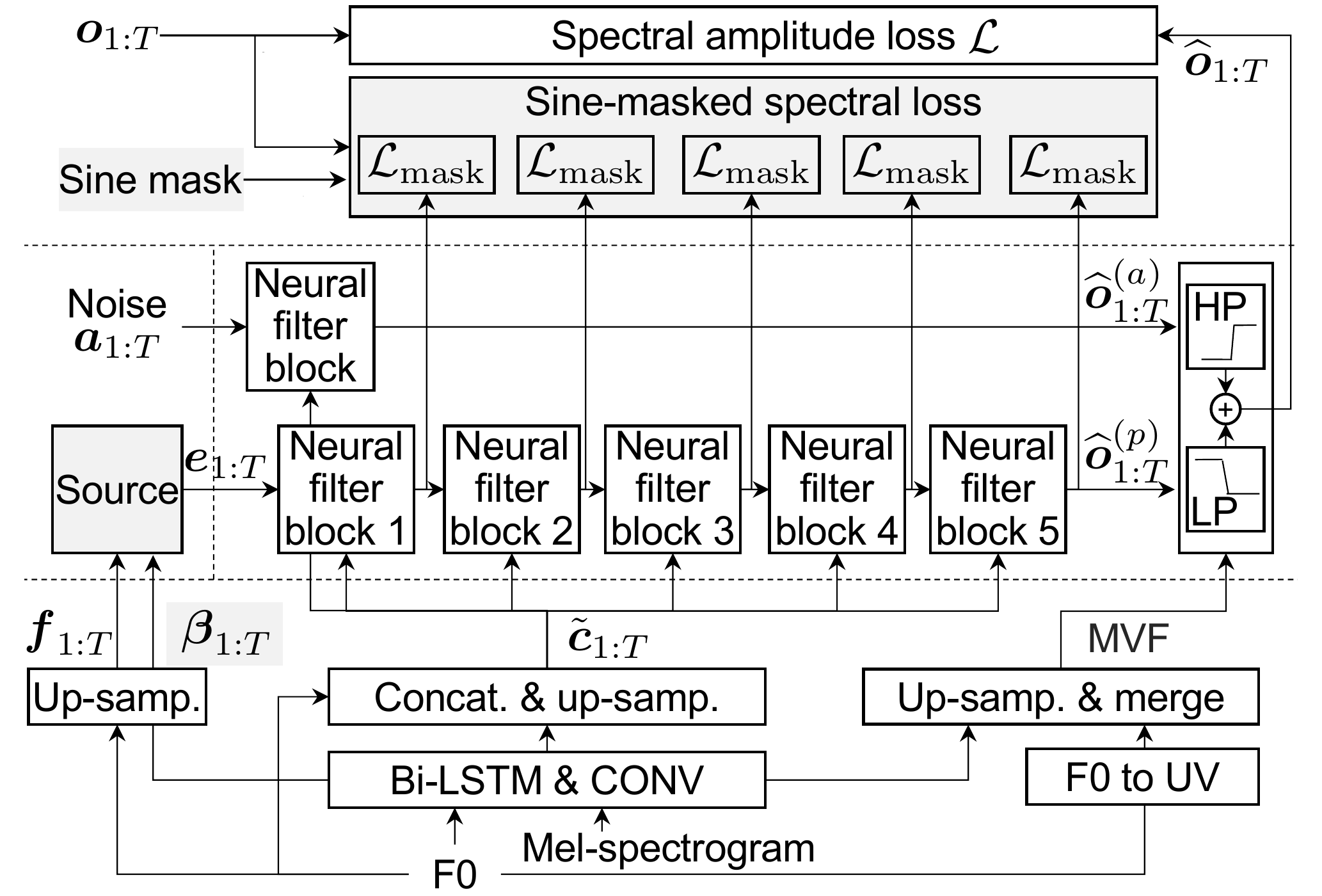}
\vspace{-6mm}
\caption{Hn-sinc-NSF during training. LP and HP denote low- and high-pass digital filters, respectively. Up-sampled feature sequence is smoothed except $\bs{f}_{1:T}$. Compared with \cite{Wang2019}, $\tilde{\bs{c}}_{1:T}$ and MVF are produced from the same Bi-LSTM and CONV layers to simplify condition module. Shaded components are those proposed in this paper. }
\label{fig:sys}
\end{figure}

Our focus in this paper is the latest harmonic-plus-noise NSF model with sinc filters (hn-sinc-NSF) \cite{Wang2019}.
In the hn-sinc-NSF model, which is shown in Figure~\ref{fig:sys}, the input frame-level F0s and Mel-spectrogram are transformed and up-sampled into feature sequences that have the same length $T$ as the target waveform $\bs{o}_{1:T}=\{o_1, \cdots, o_T\}$.  
The up-sampled F0 $\bs{f}_{1:T}=\{f_1, \cdots, f_T\}$ is fed to the source module to produce a source signal $\bs{e}_{1:T}$, and the processed acoustic feature $\tilde{\bs{c}}_{1:T}$ is fed to neural filter blocks that consist of dilated CONV layers and skip connections.
The neural filter blocks form two branches: a harmonic branch that transforms $\bs{e}_{1:T}$ into $\widehat{\bs{o}}_{1:T}^{(p)}$ and a noise branch that converts a noise sequence $\bs{a}_{1:T}$ into $\widehat{\bs{o}}_{1:T}^{(a)}$. 
The two signals $\widehat{\bs{o}}_{1:T}^{(p)}$ and $\widehat{\bs{o}}_{1:T}^{(a)}$ are then filtered by low-pass and high-pass sinc filters, respectively, and summed as the output waveform $\widehat{\bs{o}}_{1:T}$. The time-variant cut-off frequency of the sinc filters, or maximum voiced frequency (MVF) \cite{stylianou2001applying}, is predicted from the input acoustic features.  The model is trained by minimizing spectral amplitude losses between $\widehat{\bs{o}}_{1:T}$ and ${\bs{o}}_{1:T}$. 

The source signal $\bs{e}_{1:T}=\{e_1, \cdots, e_T\}$ in the harmonic branch is based on sine waveforms and can be generated in two steps. First, with the up-sampled F0 $\bs{f}_{1:T}$, a sine waveform $\bs{e}_{1:T}^{<h>}$ that carries the F0 or the $h$-th harmonic is generated by
\begin{align}
{e}_t^{<h>} = \begin{dcases}
\alpha\sin(\sum_{k=1}^{t}2\pi\frac{h{f}_k}{N_s} + {\phi}) + {n}_t, &\text{if }{f_t}>0 \\ 
\frac{\alpha}{3\sigma} {n}_t, & \text{if } f_t = 0\\ 
\end{dcases},
\label{eq:sine}
\end{align}
where $f_t=0$ denotes being unvoiced at the $t$-th time step, $\phi\in[-\pi, \pi]$ is a random initial phase, $N_s$ is the waveform sampling rate, and ${n}_t \sim \mathcal{N}(0, \sigma^2)$ is a Gaussian noise. 
Then, for each $t\in\{1,\cdots, T\}$,  ${e}_{t}$ is computed as
\begin{equation}
{{e}_{t}}=\text{tanh}(\sum_{h=1}^{H}w_h{{e}}_{t}^{<h>} + w_b),
\label{eq:tanh}
\end{equation} 
where $H$ is the number of harmonics and $\{w_1, \cdots w_H, w_b\}$ are the weights of a trainable feedforward layer with a tanh activation function. 
In our implementation, we used $H=8$, $\sigma=0.003$, and $\alpha=0.1$ as the hyper-parameters \cite{wangNSFall}. 
Note that $n_t$ in Eq.(\ref{eq:sine}) is independent of the noise source signal ${a}_{t}$ for the noise branch of the hn-sinc-NSF, even though both of them can be sampled from the same Gaussian distribution.

\section{Proposed methods}
\label{seq:proposed}

\begin{figure}[!t]
\centering
{\includegraphics[width=0.9\columnwidth]{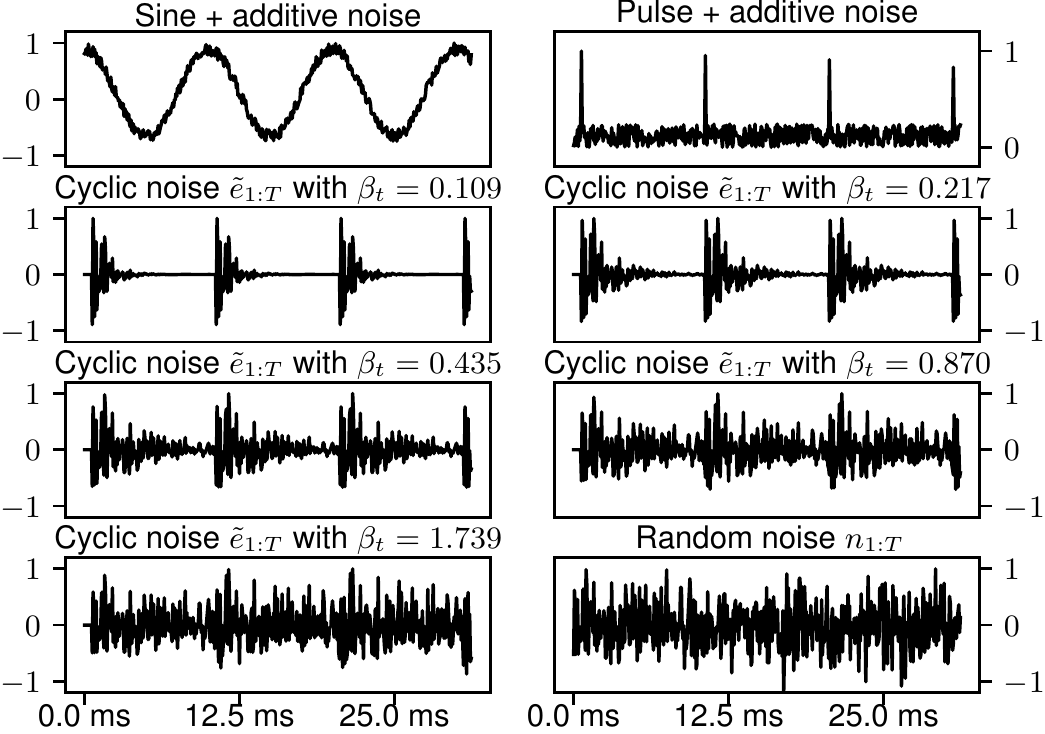}
}
\vspace{-3mm}
\caption{Example of different source signals. Maximum signal amplitude is scaled to 1 for visualization.}
\label{fig:excit}
\end{figure}

\subsection{Cyclic noise-based source signal}
Although a sine-based source is useful for generating periodic waveforms \cite{wangNSFall}, the sine shape may cause perceptual artifacts. For example, it may be inappropriate for creaky voices with irregular glottal pulses \cite{gordon2001phonation, drugman2014data} and low-pitch voices where the shape of the source signal has a greater perceptual effect \cite{rosenberg1971effect,raitio2016phase}. 
We hypothesize that a better NSF source signal for voiced sounds may contain a certain degree of randomness in the short term while preserving long-term periodicity. 
Although source signals for classical speech vocoders \cite{cabral2007towards,maia2007excitation,drugman2011deterministic,raitio2011hmm} may be used, we focus on source signals that have a simple parametric form and require no additional analysis loop.

Inspired by an experimental study on speech perception \cite{raitio2016phase}, we propose a source signal called \textit{cyclic noise}, which has a simple parametric form and can preserve both periodicity and randomness. 
Let $\bs{p}_{1:T}=[p_1, \cdots, p_T]$ denote a unit impulse train corresponding to the up-sampled F0 $\bs{f}_{1:T}$, and let $\bs{n}_{1:T}$ denote a Gaussian noise sequence. The proposed source signal $e_t$ at the $t$-th time step is computed in two steps:
\begin{align}
\tilde{e}_t = \begin{dcases}
\sum_{k=0}^{t-1} n_{k+1} \exp(-\frac{k f_t}{\beta_t N_s}) p_{t-k}, &\text{if }{f_t}>0 \\ 
{n}_t, & \text{if } f_t = 0\\ 
\end{dcases},
\label{eq:pn_definition}
\end{align}
and
\begin{equation}
{{e}_{t}}=\text{tanh}(w_1{\tilde{e}}_{t} + w_b).
\label{eq:tanh2}
\end{equation} 
Equation~(\ref{eq:tanh2}) is similar to Eq.~(\ref{eq:tanh}) but uses only $\tilde{e}_t$ rather than multiple signals with higher harmonics. 
While $\tilde{e}_t$ in the unvoiced regions is still random noise,  $\tilde{e}_t$ in the voiced regions is the convolution of $\bs{p}_{1:T}$ and an exponentially decaying noise whose value is $n_{k+1} \exp(-\frac{k f_t}{\beta_t N_s}), \forall{k}\in[0, t-1]$.  The factor $\frac{f_t}{N_s}$ makes the decaying rate independent of the period length or F0 --- no matter how long the interval between two pulses is, the noise value is scaled by $\exp(-\frac{1}{\beta_t})$ when $k$ is equal to the period length $\frac{N_s}{f_t}$. In such a manner, $\beta_t$ becomes the main parameter that controls the decaying rate.

By adjusting $\beta_t$, we can change the randomness and periodicity of the cyclic noise-based source signal for voiced sounds. Figure~\ref{fig:excit} illustrates a few source signals with different values of $\beta_t$ for $t\in[1, T]$. Note that when $\beta_t=0.435$, the noise amplitude at $t=\frac{N_s}{f_t}$ is scaled by $\exp(-\frac{1}{0.435})\approx{0.1}$. When $\beta_t$ is smaller, the random noise decays more quickly, and the source signal is more similar to the pulse train. When $\beta_t$ is increased, the source signal becomes noisy and less periodic.

We produce $\bs{p}_{1:T}$ by setting $p_t=1$, where $t$ corresponds to the local maxima of the sine waveform that carries the F0. Otherwise, $p_t = 0$. The noise sequence $\bs{n}_{1:T}$ is drawn from the same Gaussian distribution as that in Eq.~(\ref{eq:sine}), and it changes for every input training sample. 


\subsection{Predicting $\beta_t$ from acoustic features}
We may treat the decaying rate $\beta$ as a hyper-parameter and decide its value through trial and error. 
Alternatively, we can use a small neural network to predict the value of $\beta_t$ from the input acoustic features for each time step $t$.

Our implementation is illustrated in Figure~\ref{fig:sys}. We use the Bi-LSTM and CONV layers in the condition module to predict a one-dimensional signal at the frame level. This signal is then up-sampled and smoothed as $\bs{\beta}_{1:T}=\{\beta_1, \cdots, \beta_T\}$. The smoothing is conducted twice by moving averaging over a window of size 320. 
During training, an $L_1$ norm $|\beta_t - 0.870|_1$ with weight 0.01 is added to prevent $\beta_t$ from going wild. The value 0.870 is used because $\beta_t = 0.870$ keeps periodicity and randomness, as shown in Figure~\ref{fig:excit}.

\subsection{Additional masked spectral loss to stabilize pitch}
In our experiments, we observed that the randomness in the cyclic noise made it more difficult to produce a waveform with stable pitch. 
Since the source signal is gradually morphed by the NSF neural filter blocks, it may be helpful if additional loss can be defined to penalize the mismatch of the harmonic structure between the natural waveform and the output of the filter blocks. 

\begin{figure}[!t]
\centering
{\includegraphics[width=\columnwidth]{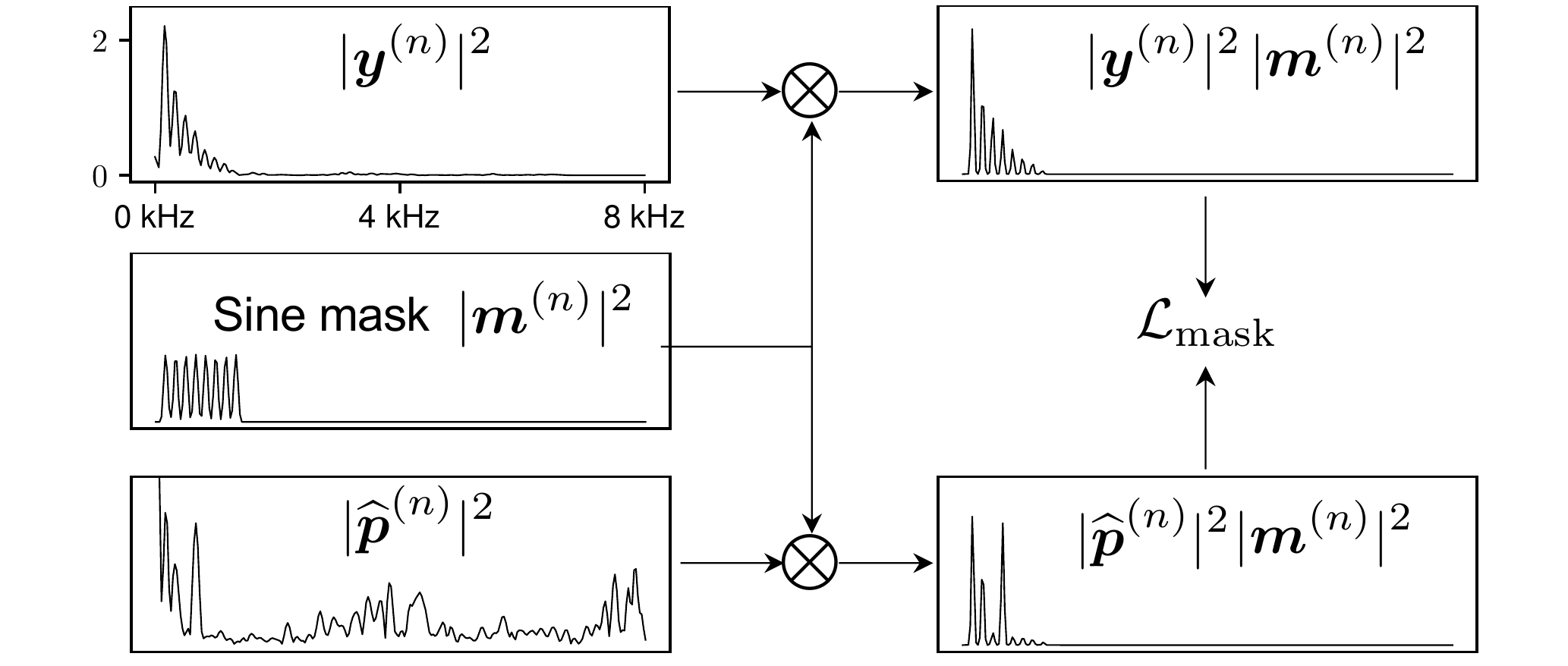}
}
\vspace{-6mm}
\caption{Sine-masked spectral loss. $\bs{y}^{(n)}$, $\widehat{\bs{p}}^{(n)}$, and $\bs{m}^{(n)}$ denote the spectrum of natural waveform, output signal from a neural filter block, and sine mask, respectively, at $n$-th frame.} 
\vspace{-2mm}
\label{fig:sine_mask}
\end{figure}

We propose a masked spectral loss for this purpose. Let ${\bs{y}}^{(n)}=[{y}^{(n)}_1, \cdots, {y}^{(n)}_K]^{\top}\in\mathbb{C}^{K}$ denote the complex-valued spectrum of the $n$-th frame of the natural waveform calculated using $K$-points FFT.
Similarly, let $\widehat{\bs{p}}^{(n)}$ and $\bs{m}^{(n)}$ denote the spectrum calculated on a neural filter block's output signal and a mask signal, respectively. 
The proposed loss is defined as
\begin{equation}
\mathcal{L}_{\text{mask}} = \frac{1}{2NK}\sum_{n=1}^{N}\sum_{k=1}^{K}\Big[\log\frac{|y_k^{(n)}|^2 |m_k^{(n)}|^2 + \eta}{|\widehat{p}_k^{(n)}|^2 |m_k^{(n)}|^2 +\eta}\Big]^2,
\label{eq:sine_mask}
\end{equation}
where $N$ is the number of frames and $\eta=1\mathrm{e}{-5}$ is a constant to ensure numerical stability. 
The mask signal is defined as $\frac{1}{H}\sum_{h=1}^{H}{\bs{e}}_{1:T}^{<h>}$, where ${\bs{e}}_{1:T}^{<h>}$ is the sine waveform that carries the $i$-th harmonics of the input F0 (see Eq.~(\ref{eq:sine})).  
The spectrum amplitude of such a mask signal, i.e., $|\bs{m}^{(n)}|^{2}$, has peaks of roughly equal height at the harmonic frequencies, and $\mathcal{L}_{\text{mask}}$ thus penalizes the mismatch of the harmonic structure rather than the whole spectral envelope, as Figure~\ref{fig:sine_mask} illustrates.

The masked spectral loss $\mathcal{L}_{\text{mask}}$ can be computed using multiple STFT configurations in the same manner as the unmasked spectral amplitude loss $\mathcal{L}$ between $\bs{o}_{1:T}$ and $\widehat{o}_{1:T}$. We used the STFT configurations in our previous work (Table III in \cite{wangNSFall}) for both $\mathcal{L}$ and $\mathcal{L}_{\text{mask}}$. Further, as Figure~\ref{fig:sys} plots, $\mathcal{L}_{\text{mask}}$ is measured between the natural waveform and the outputs of the five neural filter blocks (i.e., block 1 to block 5 in Figure~\ref{fig:sys}). 

\section{Experiments}
\label{seq:exp}

\subsection{Data and feature configuration}
\label{sec:data}
We used the data of four US English speakers from the CMU ARCTIC database \cite{kominek2004cmu} for our experiments: SLT (female), CLB (female), BDL (male), and RMS (male). Each speaker provided 1000, 64, and 64 utterances for the training, validation, and test sets, respectively, and the data were pooled together to train the experimental models in a speaker-independent manner. 

The acoustic features input to the experimental models included the Mel-spectrogram of 80 dimensions and the F0 sequence, both of which were extracted with a frame shift of 5 ms. The target speech waveforms had a sampling rate of 16 kHz and were equalized to -26 dBov using SV56 \cite{itu2011p}.  During the test stage, the natural acoustic features were used to produce the waveforms. Speaker vectors were not used.

\subsection{Experimental models}
\label{sec:models}
We utilized the hn-sinc-NSF models listed in Table~\ref{tab:sys}. 
Except the source signal and $\mathcal{L}_{\text{mask}}$, all these models used the network architecture detailed in our previous work (i.e., \texttt{sinc1-h-NSF} in \cite{Wang2019}):
five and one neural filter blocks in the harmonic and noise branches, respectively, ten dilated-CONV layers in each block, and sinc filters of order $31$. The $k$-th dilated-CONV layer in each neural filter block had a dilation size of $2^{k-1}$. 
The experiments included the natural waveforms (\texttt{Natural}) and WaveNet (\texttt{WaveNet}) as a reference. 
\texttt{WaveNet} used 30 dilated CONV layers, where the $k$-layer had a dilation size of $2^{\text{mod}(k-1, 10)}$. It used the Mel-spectrogram as the input features and generated 10-bit $\mu$-law quantized waveform values. 

All the experimental models were trained for 100 epochs using an ADAM optimizer with a learning rate $0.0003$, $\beta_1=0.9$, $\beta_2=0.999$, and $\epsilon=10^{-8}$ \cite{kingma2014adam}. The model parameters from the epoch with the lowest validation error were saved. 

\subsection{Results and analysis}
\label{sec:results}

Each trained experimental model generated 264 test set waveforms (64 utterances per speaker). The quality of the generated waveforms was evaluated by means of a large-scale crowd-sourced listening test. All the waveforms were converted to a 16-bit PCM format for the test\footnote{Code, models, and samples: \url{https://nii-yamagishilab.github.io/samples-nsf/nsf-v4.html}.}.

\begin{figure}[!t]
    \centering
    \begin{subfigure}[b]{\columnwidth}
        \includegraphics[width=\columnwidth]{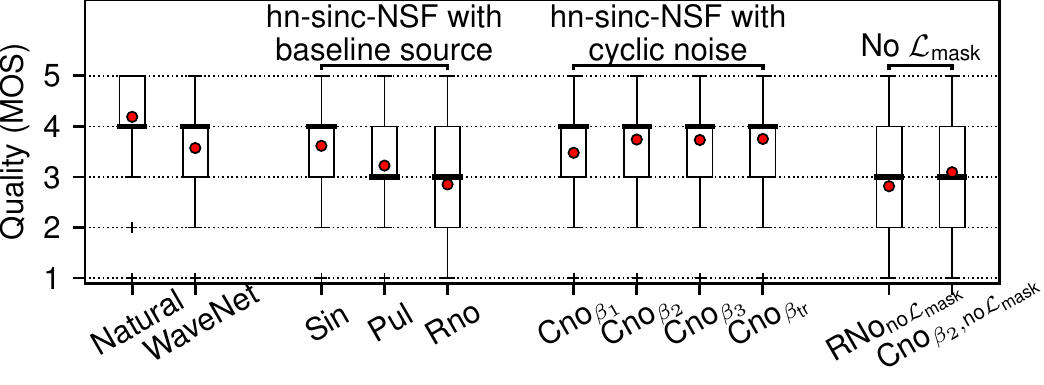}
        \vspace{-8mm}
        \caption{Results of all speakers pooled together}
        \label{fig:fig_mos_all}
    \end{subfigure}
    \quad
    
    \begin{subfigure}[b]{\columnwidth}
        \includegraphics[width=\columnwidth]{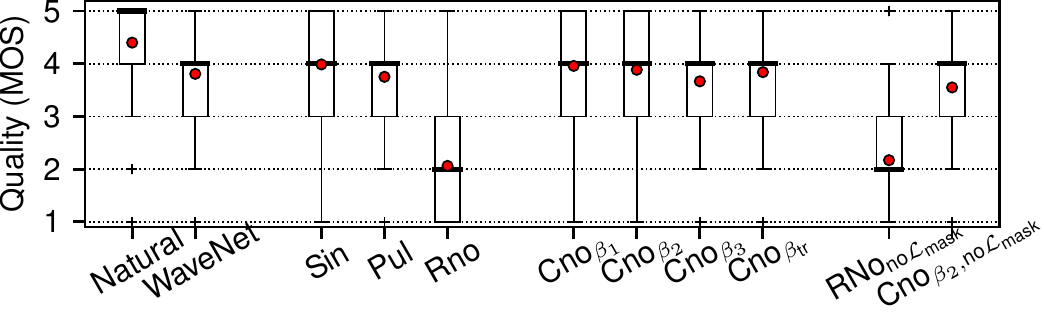}
        \vspace{-8mm}
        \caption{Results on SLT (female speaker)}
        \label{fig:fig_mos_slt}
    \end{subfigure}
    \quad
    
    \begin{subfigure}[b]{\columnwidth}
        \includegraphics[width=\columnwidth]{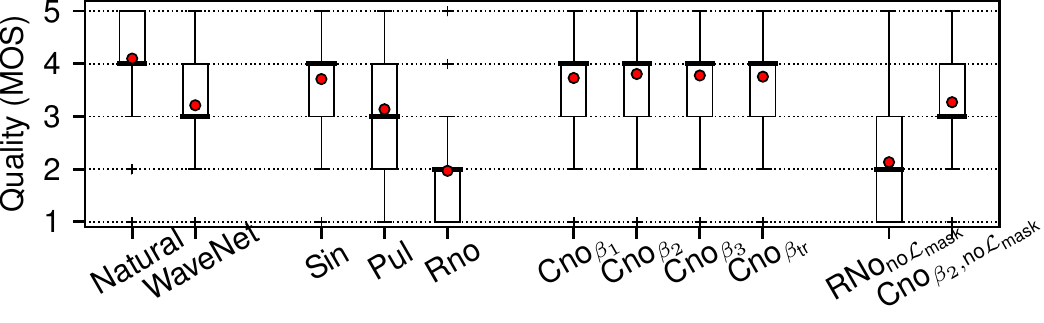}
        \vspace{-8mm}
        \caption{Results on CLB (female speaker)}
        \label{fig:fig_mos_clb}
    \end{subfigure}
    \quad
    
    \begin{subfigure}[b]{\columnwidth}
        \includegraphics[width=\columnwidth]{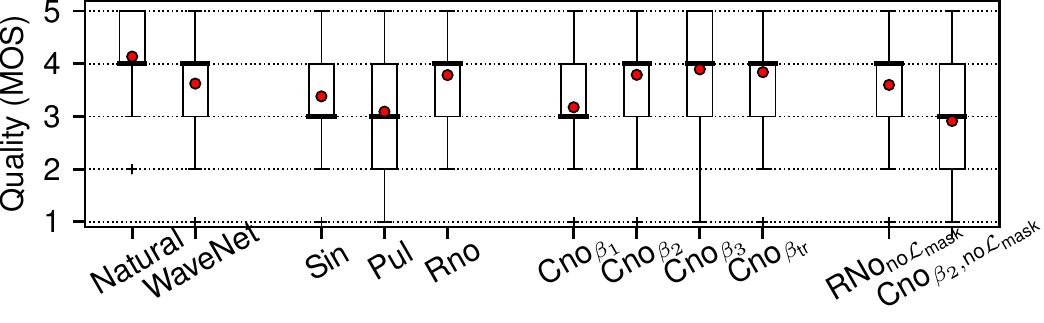}
        \vspace{-8mm}
        \caption{Results on RMS (male speaker)}
        \label{fig:fig_mos_rms}
    \end{subfigure}
    \quad
    
    \begin{subfigure}[b]{\columnwidth}
        \includegraphics[width=\columnwidth]{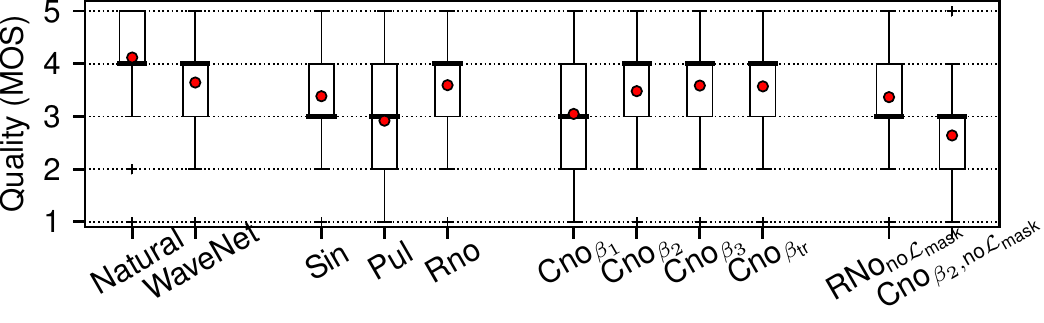}
        \vspace{-8mm}
        \caption{Results on BDL (male speaker)}
        \label{fig:fig_mos_bdl}
    \end{subfigure}
 \vspace{-4mm}
\caption{Boxplots of MOS. Dot denotes mean value of MOS. }
\vspace{-6mm}
\label{fig:mos}
\end{figure}

Each test round, an evaluator was asked to complete 22 test pages. The evaluator listened to one waveform on each page and rated its quality on a 1-to-5 mean opinion score (MOS) scale. The 22 test pages contained two randomly selected natural waveforms and two generated waveforms from each of the experimental models from the same speaker. The page order was randomly shuffled, and the waveform on each page could be played multiple times. 
Around 710 evaluators participated in the listening test and 792 test rounds were conducted. Accordingly, each experimental model received 1584 MOS scores, the boxplots of which are shown in Figure~\ref{fig:mos}. 

The results reveal interesting cross-gender differences. As the scores of \texttt{Sin}, \texttt{Pul}, and \texttt{Rno} demonstrate, the sine and pulse were better than the random noise for female speakers while the random noise outperformed the sine and pulse for male speakers. The differences in both cases were statistically significant (Mann-Whitney two-sided test, $p<0.01$)\footnote{Results of significant tests are uploaded to the website above.}.  
While cross-gender differences have also been observed on the source signals for classical speech vocoders \cite{raitio2016phase, drugman2014excitation, skoglund1997audibility}, new findings can be observed here.
For male voices, for example, \texttt{Rno} produced periodic voiced sounds, and they were perceptually better than those using sine or pulse. The performance of \texttt{Rno} suggests that the neural filter blocks can transform random noise into periodic waveforms with long periodic cycles, which is difficult for linear filters in classical vocoders. 
For the female voice, we found that it was still difficult for neural filter blocks to produce periodic waveforms with a short period cycle, so it is better to use periodic source signals.

\begin{figure}[!t]
\centering
{\includegraphics[width=\columnwidth]{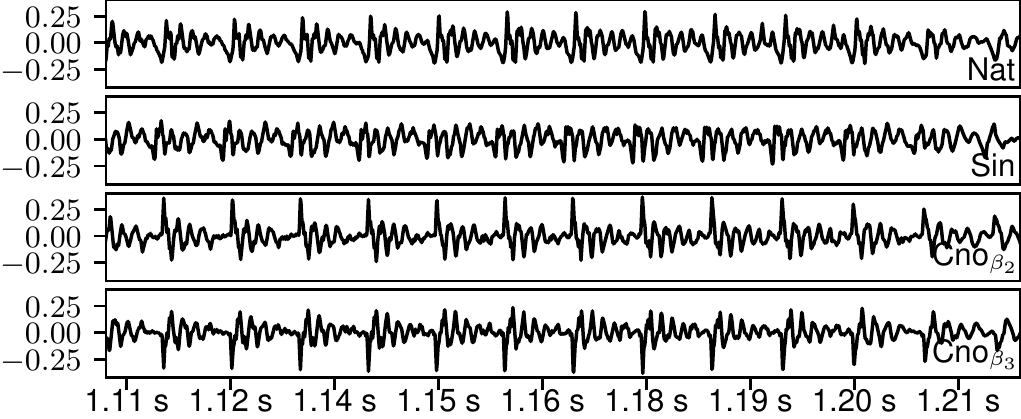}
}
\vspace{-6mm}
\caption{Waveform sample from BDL (arctic\_b0477).} 
\vspace{-6mm}
\label{fig:wave}
\end{figure}

The cyclic noise worked well for both the female and male voices when a good $\beta$ was selected. Both \texttt{Cno}${}_{\beta_2}$ and \texttt{Cno}${}_{\beta_3}$ performed quite well: they outperformed \texttt{Sin} on the male voices and were as good as \texttt{Sin} on the female voices. 
For female speakers, the cyclic noises in \texttt{Cno}${}_{\beta_2}$ and \texttt{Cno}${}_{\beta_3}$ were quasi-periodic (as Figure~\ref{fig:excit} illustrates) and may be sufficient for generating periodic waveforms. For male speakers, \texttt{Cno}${}_{\beta_2}$ and \texttt{Cno}${}_{\beta_3}$ were perceptually better than \texttt{Sin} because the generated waveforms in each periodic cycle had one obvious spike in the time domain, which is similar to how natural waveforms behave (as Figure~\ref{fig:wave} shows). In contrast, the waveform from \texttt{Sin} scattered within each periodic cycle in the time domain. \texttt{Cno}${}_{\beta_1}$ performed worse on male voices and its generated speech was perceptually similar to those from \texttt{Sin}. 

Interestingly, compared with the LPC analysis on low pitch voices (e.g., M4 in \cite{raitio2016phase}),  the waveforms from \texttt{Cno}${}_{\beta_2}$ and \texttt{Cno}${}_{\beta_3}$ were perceptually similar to those of using a zero-phase source signal in LPC, while the waveforms from \texttt{Sin} sounded similar to the LPC-produced waveforms using cyclostationary random phase source signals, even though the source signals in \texttt{Cno}${}_{\beta_2}$ and \texttt{Cno}${}_{\beta_3}$ were cyclostationary rather than zero-phased. The discrepancy from the LPC analysis again may be due to the difference between NNs and linear filters in LPC.

Although we used \texttt{Cno}${}_{\beta_{\text{tr}}}$ to infer time-variant and speaker-dependent $\beta_t$, the predicted $\beta_t$ was roughly equal to $\beta=0.86$ for all $t$ and all speakers. This may be due to the difficulty of optimization. Another reason could be that a better spectral loss is necessary to evaluate the difference caused by the varying $\beta$.

\begin{table}[!t]
\caption{Experimental models based on hn-sinc-NSF.}
\vspace{-6mm}
\begin{center}
\addtolength{\tabcolsep}{-2pt} 
{\begin{tabular}{llc}
\hline\hline
Model & Type of source signal & $\mathcal{L}_{mask}$  \\
\hline
\texttt{Sin}  & Sine-based waveform (Eq.~(\ref{eq:sine}-\ref{eq:tanh})) & -- \\
\texttt{Pul}  & Pulse with noise (Gaussian) & -- \\
\texttt{Rno} & Random noise (Gaussian) & + \\
\hline
\texttt{Cno}${}_{\beta_{1}}$ & Cyclic noise, $\beta_t=0.435$ & + \\
\texttt{Cno}${}_{\beta_{2}}$ & Cyclic noise, $\beta_t=0.870$ & + \\
\texttt{Cno}${}_{\beta_{3}}$ & Cyclic noise, $\beta_t=1.739$ & + \\
\texttt{Cno}${}_{\beta_{\text{tr}}}$ & Cyclic noise,  trainable $\beta_t$ & + \\
\hline
\texttt{Rno}${}_{\text{no}\mathcal{L}_{\text{mask}}}$ &  Random Gaussian noise & -- \\
\texttt{Cno}${}_{\beta_{2}\text{no}\mathcal{L}_{\text{mask}}}$ &  Cyclic noise,  $\beta_t=0.870$ & -- \\
\hline
\end{tabular}}
\vspace{-4mm}
\end{center}
\label{tab:sys}
\end{table}%

As a final remark, \texttt{Cno}${}_{\beta_{2}}$ outperformed \texttt{Cno}${}_{\beta_{2}\text{no}\mathcal{L}_{\text{mask}}}$ for all four speakers, and the difference was statistically significant. This result indicates the  usefulness of the masked spectral loss $\mathcal{L}_{\text{mask}}$ for \texttt{Cno}${}_{\beta_{\text{2}}}$. 
However, $\mathcal{L}_{\text{mask}}$ was less effective for \texttt{Rno}, especially on female speakers. We conclude that $\mathcal{L}_{\text{mask}}$ is insufficient to guide the model to reproduce stable pitch from noise source signals for female speakers.

\section{Conclusion}
\label{sec:conclude}
We proposed a new type of source signal called cyclic noise for NSF models. Cyclic noise has a simple parametric form based on the convolution of a pulse train and an exponentially decaying noise sequence. We demonstrated through experiments that an appropriate $\beta$ can be specified to maintain the periodicity and introduce randomness to the cyclic noise signal. Compared with the sine-based source signal, which worked well for female speakers but not for male speakers, the proposed cyclic noise worked well for both female and male voices. 

Although the findings in this paper were derived from NSF models, they showed interesting differences from those on the classical vocoders. Future work will examine the link between the perceptual difference and the phase \cite{raitio2016phase,skoglund1997audibility, pobloth1999phase}.

\noindent
\textbf{Acknowledgement:}
This work was partially supported by a JST CREST Grant (JPMJCR18A6, VoicePersonae project), Japan, MEXT KAKENHI Grants (19K24371, 16H06302, 17H04687, 18H04120, 18H04112, 18KT0051), Japan, and Google AI for Japan program. The experiments partially were conducted on TSUBAME 3.0 of Tokyo Institute of Technology. We thank Mr. Yusuke Yasuda for setting up the server for listening test. We also thank Dr. Heiga Zen and Dr. Yotaro Kubo for comments and discussions. 

\bibliographystyle{IEEEtran}

\bibliography{../../../../PAPER/library.bib}

\appendix
\onecolumn
\section{Results from Mann-Whitney U test}
In the tables below, $\square$ denotes significant difference ($p<0.01$), and $\bullet$ denotes insignificant difference.
\begin{table}[h]
\caption{Results over all speakers}
\vspace{-5mm}
\begin{center}
\begin{tabular}{|c|c|c|c|c|c|c|c|c|c|c|c|}
\hline 
 & Nat  & WaN  & Sin  & Pul  & Rno  & Cno${}_{\beta_{1}}$  & Cno${}_{\beta_{2}}$  & Cno${}_{\beta_{3}}$  & Cno${}_{\beta_{\text{tr}}}$  & RNo${}_{\text{no} \mathcal{L}_{\text{mask}}}$  & Cno${}_{\beta_{\text{tr}}, \text{no} \mathcal{L}_{\text{mask}}}$ \\ \hline
Nat & - &  &  &  &  &  &  &  &  &  & \\ \hline
WaN & $\bullet$  & - &  &  &  &  &  &  &  &  & \\ \hline
Sin & $\bullet$  & $\square$  & - &  &  &  &  &  &  &  & \\ \hline
Pul & $\bullet$  & $\bullet$  & $\bullet$  & - &  &  &  &  &  &  & \\ \hline
Rno & $\bullet$  & $\bullet$  & $\bullet$  & $\bullet$  & - &  &  &  &  &  & \\ \hline
Cno${}_{\beta_{1}}$ & $\bullet$  & $\bullet$  & $\bullet$  & $\bullet$  & $\bullet$  & - &  &  &  &  & \\ \hline
Cno${}_{\beta_{2}}$ & $\bullet$  & $\bullet$  & $\bullet$  & $\bullet$  & $\bullet$  & $\bullet$  & - &  &  &  & \\ \hline
Cno${}_{\beta_{3}}$ & $\bullet$  & $\bullet$  & $\bullet$  & $\bullet$  & $\bullet$  & $\bullet$  & $\square$  & - &  &  & \\ \hline
Cno${}_{\beta_{\text{tr}}}$ & $\bullet$  & $\bullet$  & $\bullet$  & $\bullet$  & $\bullet$  & $\bullet$  & $\square$  & $\square$  & - &  & \\ \hline
RNo${}_{\text{no} \mathcal{L}_{\text{mask}}}$ & $\bullet$  & $\bullet$  & $\bullet$  & $\bullet$  & $\square$  & $\bullet$  & $\bullet$  & $\bullet$  & $\bullet$  & - & \\ \hline
Cno${}_{\beta_{\text{tr}}, \text{no} \mathcal{L}_{\text{mask}}}$ & $\bullet$  & $\bullet$  & $\bullet$  & $\bullet$  & $\bullet$  & $\bullet$  & $\bullet$  & $\bullet$  & $\bullet$  & $\bullet$  & -\\ \hline
\end{tabular}
\end{center}
\label{tab:mos_all}
\end{table}%

\begin{table}[h]
\caption{Results on SLT}
\vspace{-5mm}
\begin{center}
\begin{tabular}{|c|c|c|c|c|c|c|c|c|c|c|c|}
\hline
 & Nat  & WaN  & Sin  & Pul  & Rno  & Cno${}_{\beta_{1}}$  & Cno${}_{\beta_{2}}$  & Cno${}_{\beta_{3}}$  & Cno${}_{\beta_{\text{tr}}}$  & RNo${}_{\text{no} \mathcal{L}_{\text{mask}}}$  & Cno${}_{\beta_{\text{tr}}, \text{no} \mathcal{L}_{\text{mask}}}$ \\ \hline
Nat & - &  &  &  &  &  &  &  &  &  & \\ \hline
WaN & $\bullet$  & - &  &  &  &  &  &  &  &  & \\ \hline
Sin & $\bullet$  & $\bullet$  & - &  &  &  &  &  &  &  & \\ \hline
Pul & $\bullet$  & $\square$  & $\bullet$  & - &  &  &  &  &  &  & \\ \hline
Rno & $\bullet$  & $\bullet$  & $\bullet$  & $\bullet$  & - &  &  &  &  &  & \\ \hline
Cno${}_{\beta_{1}}$ & $\bullet$  & $\square$  & $\square$  & $\bullet$  & $\bullet$  & - &  &  &  &  & \\ \hline
Cno${}_{\beta_{2}}$ & $\bullet$  & $\square$  & $\square$  & $\square$  & $\bullet$  & $\square$  & - &  &  &  & \\ \hline
Cno${}_{\beta_{3}}$ & $\bullet$  & $\square$  & $\bullet$  & $\square$  & $\bullet$  & $\bullet$  & $\bullet$  & - &  &  & \\ \hline
Cno${}_{\beta_{\text{tr}}}$ & $\bullet$  & $\square$  & $\square$  & $\square$  & $\bullet$  & $\square$  & $\square$  & $\bullet$  & - &  & \\ \hline
RNo${}_{\text{no} \mathcal{L}_{\text{mask}}}$ & $\bullet$  & $\bullet$  & $\bullet$  & $\bullet$  & $\square$  & $\bullet$  & $\bullet$  & $\bullet$  & $\bullet$  & - & \\ \hline
Cno${}_{\beta_{\text{tr}}, \text{no} \mathcal{L}_{\text{mask}}}$ & $\bullet$  & $\bullet$  & $\bullet$  & $\square$  & $\bullet$  & $\bullet$  & $\bullet$  & $\square$  & $\bullet$  & $\bullet$  & -\\ \hline
\end{tabular}
\end{center}
\label{tab:mos_slt}
\end{table}%

\begin{table}[h]
\caption{Results on CLB}
\vspace{-5mm}
\begin{center}
\begin{tabular}{|c|c|c|c|c|c|c|c|c|c|c|c|}
\hline
 & Nat  & WaN  & Sin  & Pul  & Rno  & Cno${}_{\beta_{1}}$  & Cno${}_{\beta_{2}}$  & Cno${}_{\beta_{3}}$  & Cno${}_{\beta_{\text{tr}}}$  & RNo${}_{\text{no} \mathcal{L}_{\text{mask}}}$  & Cno${}_{\beta_{\text{tr}}, \text{no} \mathcal{L}_{\text{mask}}}$ \\ \hline
Nat & - &  &  &  &  &  &  &  &  &  & \\ \hline
WaN & $\bullet$  & - &  &  &  &  &  &  &  &  & \\ \hline
Sin & $\bullet$  & $\bullet$  & - &  &  &  &  &  &  &  & \\ \hline
Pul & $\bullet$  & $\square$  & $\bullet$  & - &  &  &  &  &  &  & \\ \hline
Rno & $\bullet$  & $\bullet$  & $\bullet$  & $\bullet$  & - &  &  &  &  &  & \\ \hline
Cno${}_{\beta_{1}}$ & $\bullet$  & $\bullet$  & $\square$  & $\bullet$  & $\bullet$  & - &  &  &  &  & \\ \hline
Cno${}_{\beta_{2}}$ & $\bullet$  & $\bullet$  & $\square$  & $\bullet$  & $\bullet$  & $\square$  & - &  &  &  & \\ \hline
Cno${}_{\beta_{3}}$ & $\bullet$  & $\bullet$  & $\square$  & $\bullet$  & $\bullet$  & $\square$  & $\square$  & - &  &  & \\ \hline
Cno${}_{\beta_{\text{tr}}}$ & $\bullet$  & $\bullet$  & $\square$  & $\bullet$  & $\bullet$  & $\square$  & $\square$  & $\square$  & - &  & \\ \hline
RNo${}_{\text{no} \mathcal{L}_{\text{mask}}}$ & $\bullet$  & $\bullet$  & $\bullet$  & $\bullet$  & $\square$  & $\bullet$  & $\bullet$  & $\bullet$  & $\bullet$  & - & \\ \hline
Cno${}_{\beta_{\text{tr}}, \text{no} \mathcal{L}_{\text{mask}}}$ & $\bullet$  & $\square$  & $\bullet$  & $\square$  & $\bullet$  & $\bullet$  & $\bullet$  & $\bullet$  & $\bullet$  & $\bullet$  & -\\ \hline
\end{tabular}
\end{center}
\label{tab:mos_clb}
\end{table}%

\begin{table}[h]
\caption{Results on RMS}
\vspace{-5mm}
\begin{center}
\begin{tabular}{|c|c|c|c|c|c|c|c|c|c|c|c|}
\hline
 & Nat  & WaN  & Sin  & Pul  & Rno  & Cno${}_{\beta_{1}}$  & Cno${}_{\beta_{2}}$  & Cno${}_{\beta_{3}}$  & Cno${}_{\beta_{\text{tr}}}$  & RNo${}_{\text{no} \mathcal{L}_{\text{mask}}}$  & Cno${}_{\beta_{\text{tr}}, \text{no} \mathcal{L}_{\text{mask}}}$ \\ \hline
Nat & - &  &  &  &  &  &  &  &  &  & \\ \hline
WaN & $\bullet$  & - &  &  &  &  &  &  &  &  & \\ \hline
Sin & $\bullet$  & $\bullet$  & - &  &  &  &  &  &  &  & \\ \hline
Pul & $\bullet$  & $\bullet$  & $\bullet$  & - &  &  &  &  &  &  & \\ \hline
Rno & $\bullet$  & $\square$  & $\bullet$  & $\bullet$  & - &  &  &  &  &  & \\ \hline
Cno${}_{\beta_{1}}$ & $\bullet$  & $\bullet$  & $\bullet$  & $\square$  & $\bullet$  & - &  &  &  &  & \\ \hline
Cno${}_{\beta_{2}}$ & $\bullet$  & $\square$  & $\bullet$  & $\bullet$  & $\square$  & $\bullet$  & - &  &  &  & \\ \hline
Cno${}_{\beta_{3}}$ & $\bullet$  & $\bullet$  & $\bullet$  & $\bullet$  & $\square$  & $\bullet$  & $\square$  & - &  &  & \\ \hline
Cno${}_{\beta_{\text{tr}}}$ & $\bullet$  & $\bullet$  & $\bullet$  & $\bullet$  & $\square$  & $\bullet$  & $\square$  & $\square$  & - &  & \\ \hline
RNo${}_{\text{no} \mathcal{L}_{\text{mask}}}$ & $\bullet$  & $\square$  & $\bullet$  & $\bullet$  & $\bullet$  & $\bullet$  & $\bullet$  & $\bullet$  & $\bullet$  & - & \\ \hline
Cno${}_{\beta_{\text{tr}}, \text{no} \mathcal{L}_{\text{mask}}}$ & $\bullet$  & $\bullet$  & $\bullet$  & $\square$  & $\bullet$  & $\bullet$  & $\bullet$  & $\bullet$  & $\bullet$  & $\bullet$  & -\\ \hline
\end{tabular}
\end{center}
\label{tab:mos_rms}
\end{table}%

\begin{table}[h]
\caption{Results on BDL}
\vspace{-5mm}
\begin{center}
\begin{tabular}{|c|c|c|c|c|c|c|c|c|c|c|c|}
\hline
 & Nat  & WaN  & Sin  & Pul  & Rno  & Cno${}_{\beta_{1}}$  & Cno${}_{\beta_{2}}$  & Cno${}_{\beta_{3}}$  & Cno${}_{\beta_{\text{tr}}}$  & RNo${}_{\text{no} \mathcal{L}_{\text{mask}}}$  & Cno${}_{\beta_{\text{tr}}, \text{no} \mathcal{L}_{\text{mask}}}$ \\ \hline
Nat & - &  &  &  &  &  &  &  &  &  & \\ \hline
WaN & $\bullet$  & - &  &  &  &  &  &  &  &  & \\ \hline
Sin & $\bullet$  & $\bullet$  & - &  &  &  &  &  &  &  & \\ \hline
Pul & $\bullet$  & $\bullet$  & $\bullet$  & - &  &  &  &  &  &  & \\ \hline
Rno & $\bullet$  & $\square$  & $\bullet$  & $\bullet$  & - &  &  &  &  &  & \\ \hline
Cno${}_{\beta_{1}}$ & $\bullet$  & $\bullet$  & $\bullet$  & $\square$  & $\bullet$  & - &  &  &  &  & \\ \hline
Cno${}_{\beta_{2}}$ & $\bullet$  & $\square$  & $\square$  & $\bullet$  & $\square$  & $\bullet$  & - &  &  &  & \\ \hline
Cno${}_{\beta_{3}}$ & $\bullet$  & $\square$  & $\bullet$  & $\bullet$  & $\square$  & $\bullet$  & $\square$  & - &  &  & \\ \hline
Cno${}_{\beta_{\text{tr}}}$ & $\bullet$  & $\square$  & $\square$  & $\bullet$  & $\square$  & $\bullet$  & $\square$  & $\square$  & - &  & \\ \hline
RNo${}_{\text{no} \mathcal{L}_{\text{mask}}}$ & $\bullet$  & $\bullet$  & $\square$  & $\bullet$  & $\bullet$  & $\bullet$  & $\square$  & $\bullet$  & $\bullet$  & - & \\ \hline
Cno${}_{\beta_{\text{tr}}, \text{no} \mathcal{L}_{\text{mask}}}$ & $\bullet$  & $\bullet$  & $\bullet$  & $\bullet$  & $\bullet$  & $\bullet$  & $\bullet$  & $\bullet$  & $\bullet$  & $\bullet$  & -\\ \hline
\end{tabular}
\end{center}
\label{tab:mos_bdl}
\end{table}%

\end{document}